\documentclass[epj]{svjour}
\usepackage{graphics}
\begin{document}
\newcommand {\be}{\begin{equation}}
\newcommand {\ee}{\end{equation}}
\newcommand {\bea}{\begin{eqnarray}}
\newcommand {\eea}{\end{eqnarray}}
\newcommand {\nn}{\nonumber}

\title{Anisotropic superconductivity in PrOs$_{4}$Sb$_{12}$}

\author{David Parker \and Kazumi Maki \and Stephan Haas}

\institute{Department of Physics and Astronomy, University of Southern
California, Los Angeles, CA 90089-0484 USA  \\ Phone (213)740-1104, e-mail
davidspa@usc.edu}

\date{Received:  / Revised version: }

\abstract{
Recently two anisotropic superconducting gap functions have been observed 
in the skutterudite PrOs$_4$Sb$_{12}$. These order parameters are spin-triplet.  There are at least 2 distinct phases
in a magnetic field, bearing some resemblance to superfluid $ ^{3}$He.  Here we
present an analysis of the thermodynamic properties in these two superconducting states
within the weak-coupling BCS theory.
\PACS{{74.20.Fg}{BCS theory and its development}}
}
\maketitle
 
\section{Introduction}

Superconductivity in the body-centered
cubic heavy-fermion (HF) skutterudite PrOs$_4$Sb$_{12}$
was discovered in 2002 
by Bauer et al\cite{bauer,vollmer,kotegawa}.  Since then many experimental and theoretical 
studies of this compound have been reported.  This compound possesses
several interesting and unusual characteristics: two distinct phases (the A phase and B phase) in a magnetic field,
nodal superconductivity with point nodes, and triplet pairing with chiral symmetry breaking 
\cite{aoki,maki1,tou}.  The phase diagram is still controversial.  In Fig. 1 
recent measurements by Measson et al \cite{measson} are shown.  

\begin{figure}
\resizebox{0.75\columnwidth}{!}{\includegraphics{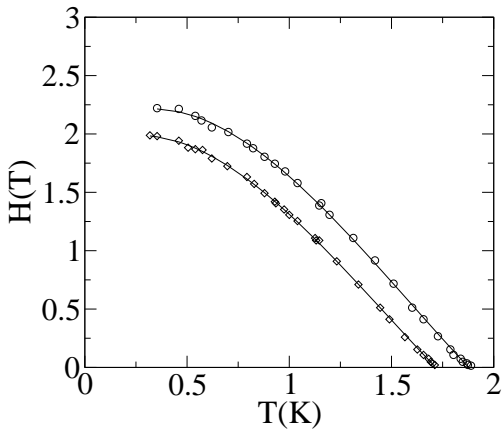}}
\caption{Phase diagram by Measson et al\cite{measson}.  The top line and points
are the upper critical field H$_{c2}$, while the lower ones are 
the phase boundary H'.}
\end{figure} 

It was recently observed that the magnetothermal conductivity data \cite{izawa,maki1} in this compound
are consistent with anisotropic superconductivity using the gap functions
\bea
{\bf \Delta}_{A}({\bf k}) &=& {\bf d} e^{\pm i\phi_{i}}\frac{3}{2}(1-\hat{k}_{x}^{4}-\hat{k}_{y}^{4}-\hat{k}_{z}^{4})).\\
{\bf \Delta}_{B}({\bf k}) &=& {\bf d} e^{\pm i\phi_{3}}(1-\hat{k}_{z}^{4})
\eea
Here $e^{\pm i\phi_{i}}$ is one of $e^{i\phi_{1}}=( \hat{k}_{y}+i \hat{k}_{z})/\sqrt{\hat{k}_{y}^{2}+\hat{k}_{z}^{2}},
e^{i\phi_{2}}=(\hat{k}_{z}+i\hat{k}_{x})/\sqrt{\hat{k}_{z}^{2}+\hat{k}_{x}^{2}},
e^{i\phi_{3}}= (\hat{k}_{x}+i\hat{k}_{y})/\sqrt{\hat{k}_{x}^{2}+\hat{k}_{y}^{2}}$. 
The factor of 3/2 ensures proper normalization of the angular dependence of the order parameter.
In Eq.(2) the nodal direction is chosen to be parallel to (001). 
\begin{figure*}
\resizebox{0.75\columnwidth}{!}{
\includegraphics{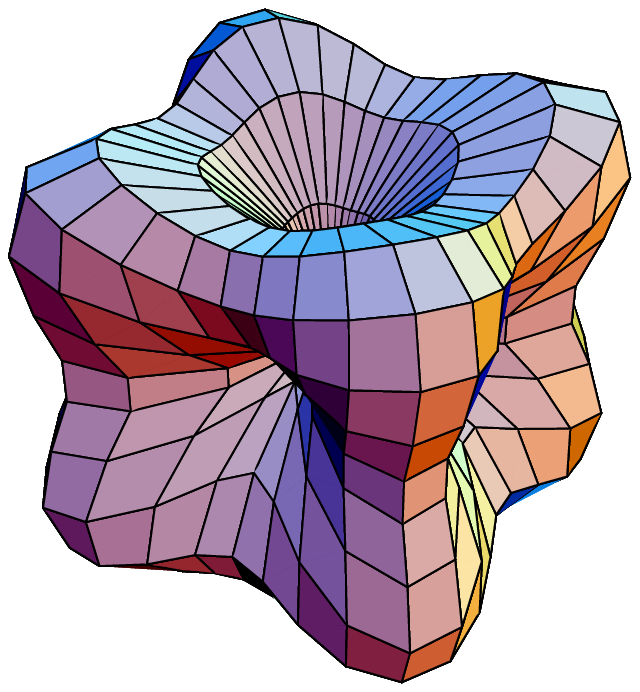}}
\resizebox{0.75\columnwidth}{!}{
\includegraphics{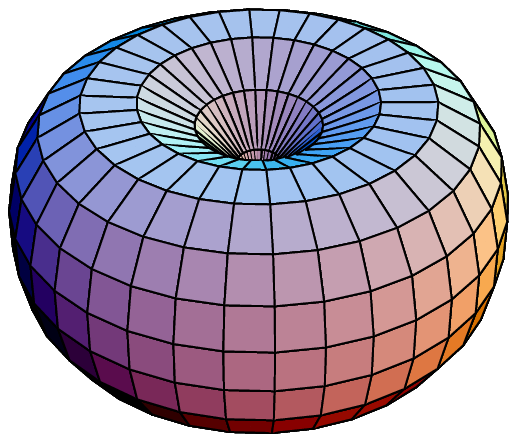}}
\caption{A-phase (left) and B-phase (right) order parameters}
\end{figure*} 

We note that the proposed order parameter (2) lies outside of the usual 
classification scheme\cite{sigrist}, 
in which order parameters correspond to a single
irreducible representation of the rotation group.  However, this hybrid order
parameter appears to be necessary to reproduce the observed B-phase 
gap structure.  A similar situation has been observed 
in the borocarbide superconductors \cite{maki3}.

The cubic symmetry of PrOs$_{4}$Sb$_{12}$ suggests order parameters 
which are invariant under the T$_{h}$ cubic tetrahedral symmetry 
group applicable to this crystal \cite{goryo}, as well as reflections 
(containing the origin) about 
the planes of the crystal parallel to the cube faces\cite{bauer2}.  
As suggested in \cite{goryo}, one
possible invariant is $\hat{k}_{x}^{2}\hat{k}_{y}^{2}+\hat{k}_{y}^{2}\hat{k}_{z}^{2}
+\hat{k}_{z}^{2}\hat{k}_{x}^{2}$.  This belongs to the A$_{1}$ representation
of T$_{h}$ \cite{goryo}.

This combination can be recast as 
$1-\hat{k}_{x}^{4}-\hat{k}_{y}^{4}-\hat{k}_{z}^{4}$, thus forming the basis 
of the proposed order parameter of the A phase. 
Furthermore, in weak-coupling BCS theory
the quasiparticle density of states and the 
thermodynamics depend only on $|f|$\cite{dahm}, 
the magnitude of the angle-dependent part of the order parameter.  For this reason, an order
parameter which breaks chiral symmetry still retains the essential features of the cubic symmetry,
and is in fact necessitated by the triplet pairing observed in this compound.\cite{aoki,maki1,tou} Triplet pairing requires that the orbital wavefunction be antisymmetric 
under particle interchange.  The phase factor proposed in Eq. (1) meets this requirement.

The proposed B-phase order parameter breaks the cubic symmetry more manifestly.  
Nevertheless, there is almost certainly a relationship between the A and B phase, 
particularly since the zero-field transition temperatures are so nearly equal.  The simplest relationship consistent with the nodes at [001] and [00-1], 
despite requiring a hybrid representation, 
would suggest $|f| \sim 1-\hat{k}_{z}^{4}$.  This order parameter 
again would have a phase factor included in $f$ to ensure
antisymmetry under interchange.  These proposed order parameters are illustrated in Fig.2 \cite{footnote}. 

We note that while the B-phase is the prevalent phase in zero magnetic field, the A-phase
exists at all temperatures below T$_{c}$ for fields between H* (the phase boundary) 
and H$_{c2}$.  

Below is a comparison of the predicted B-phase DOS with scanning 
tunneling microscopy (STM) data taken by Suderow et al \cite{suderow} at 
T=0.19 K.  The predicted B-phase DOS differs somewhat
from that presented in \cite{maki1} due to the use of the angular-dependent
quasiparticle density-of-states, as well as an accounting for the energy
and directional resolution of the STM .  Here we have assumed that the STM 
performed measurements along the nodal directions, where $\alpha$ is the
size of the momentum cone within the STM's spatial resolution.  We note that 
the nodal structure can easily be masked by performing 
STM along a limited number
of directions of single crystals.  Here we have
assumed $\Delta$ to take the B-phase weak-coupling value of 3.3 K.  We note 
that the observed small DOS for $ E < \Delta/3$ can be reproduced by choosing
$\alpha$ to be 3.0.  We observe fair agreement, with some differences apparent
surrounding the quasiparticle peak at $E=\Delta$.   
\begin{figure}
\resizebox{0.91\columnwidth}{!}{
\includegraphics{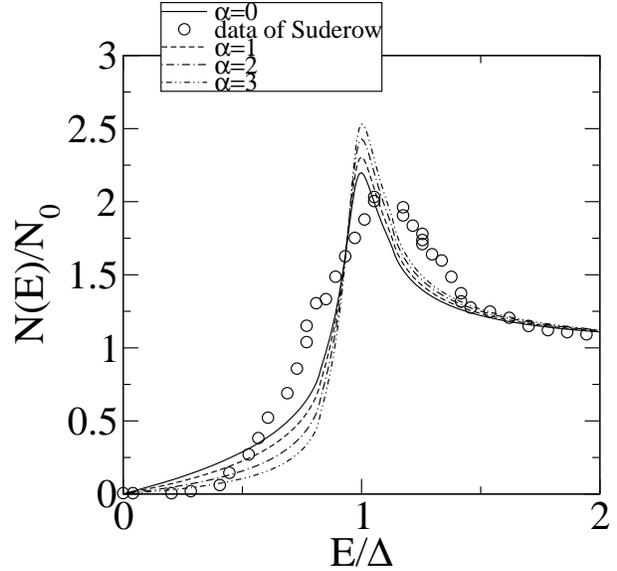}}
\caption{Comparison of quasiparticle DOS in B phase with STM data of 
Suderow et al. \cite{suderow} }
\end{figure}
In the following we analyze both phases over the entire temperature
range from $T=0$ to $T_{c}$.

\section{Weak-coupling BCS Theory}

We focus on the superconductivity in the A and B-phases of
PrOs$_{4}$Sb$_{12}$, using the $\Delta({\bf k})$ given
by Eqs. 1 and 2 with $|{\bf d}|=\Delta(T)$.

Then, within the weak-coupling theory the gap equation is given by 
\bea
\lambda^{-1}& = & 2\pi T<f^{2}>^{-1}\sum_{n}^{}\langle\frac{f^{2}}
{\sqrt{\omega_{n}^{2}+\Delta^{2}f^{2}}}\rangle\\          
& = & <f^{2}>^{-1}\int_{0}^{E_{0}}dE
\langle\frac{f^{2}}{\sqrt{E^{2}-\Delta^{2}f^{2}}}\rangle\tanh(\frac{E}{2T})
\eea
where $\lambda$ is a dimensionless coupling constant, E$_{0}$ is the cut-off energy, and
$\omega_{n}$ is the Matsubara frequency.  Here, for the A-phase $f=\frac{3}{2}
(1-\hat{k}_{x}^{4}-\hat{k}_{y}^{4} -\hat{k}_{z}^{4})$, and 
$\langle \dots \rangle$ denotes $\int d\Omega/4\pi$.
For the B-phase, $f=1-z^{4},$ and $\langle \ldots \rangle$
denotes $\int_{0}^{1}dz \ldots$.  The frequency sum in Eq. 3
is cut off at $\omega_{n}=E_{0}$.  

In the vicinity of $T=T_{c}$ and $T=0 K$, Eq. 3 and Eq. 4 can be solved
analytically.  For $T \rightarrow T_{c}$ we obtain
\bea
T_{c}=\frac{2\gamma}{\pi}E_{0}e^{-1/\lambda} &=& 1.136E_{0}e^{-1/\lambda} \\
\Delta^{2}(T) &\approx& \frac{2<f^{2}>(2\pi T_{c})^{2}\ln(T/T_{c})}{7\zeta(3) <f^{4}>} 
\eea
where $\gamma=1.78\ldots$ is the Euler constant and
\bea
\Delta(0)/T_{c} &=& \frac{\pi}{\gamma}\exp[-<f^{2}>^{-1}<f^{2}\ln(f)>]\\
                &=& 2.364, \mathrm{A-phase} \\
                &=& 1.938, \mathrm{B-phase}
\eea
In the low-temperature regime $T/\Delta(0) \ll 1$ one obtains
\bea
\ln(\Delta(T)/ \Delta(0)) = -\frac{7 \pi \zeta(3)}{8}(\frac{T}{\Delta(0)})^{3} \,\,\,\,\, 
\mathrm{A-phase} \\
\ln(\Delta(T)/ \Delta(0)) = -\frac{135 \pi \zeta(3)}{512}(\frac{T}{\Delta(0)})^{3}\,\,\,\,\,\mathrm{B-phase}
\eea
In Fig. 4 numerical solutions of $\Delta(T)/\Delta(0)$ are shown for both phases
over the entire temperature range.  
\begin{figure}
\resizebox{0.75\columnwidth}{!}{
\includegraphics{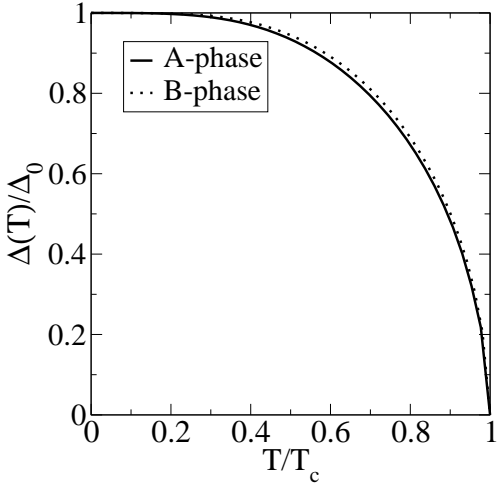}}
\caption{$\Delta(T)$ for the A and B phases}
\end{figure} 

The values of $\Delta(0)_{A}$ and $\Delta(0)_{B}$ obtained from the weak-coupling theory 
offer a possible explanation for the multiphase diagram of PrOs$_{4}$Sb$_{12}$.
The
value of the condensation energy at $T=0$ is given by
\bea
E_{0}= -\frac{1}{2}<|f|^{2}>N_{0}\Delta^{2}_{0}
\eea
where $f$ is the angular-dependent part of the order parameter and $N_{0}$ the normal state 
density of states at the Fermi level.  For the A-phase, $<|f|^{2}>$ = 3/7, whereas for the B-phase,
$<|f|^{2}>$ = 32/45.  This yields two distinct condensation energies:
\bea
E_{0}^{A} &=& -1.17 N_{0} (T_{c}^{A})^{2}, \\
E_{0}^{B} &=& -1.32 N_{0} (T_{c}^{B})^{2}.
\eea
If one uses the experimental values \cite{measson} for $T_{c}^{A}$ and $T_{c}^{B}$ as
1.887 K and 1.716 K, respectively, one finds that E$_{0}^{A}$ is slightly lower than
E$_{0}^{B}$, contrary to observation.  On the other hand, if we assume $T_{c}^{A}$ = $T_{c}^{B}$,
which the simplest interaction would give, we find  $E_{0}^{B} < E_{0}^{A}$.  
The difference between our assumed and the 
measured T$_{c}$ would in this case be
due to some unknown external perturbation, such as the effect of the 
crystalline electric field (CEF), not accounted for in this treatment.  One
possibility is that the CEF affects superconductivity in 
the two phases differently, resulting in a
difference in measured transition temperatures.
Indeed, if we assume that, absent such an effect, we would have
T$_{c}^{A}$ = T$_{c}^{B}$ we find $\Omega_{0}^{B} < \Omega_{0}^{A}$ for all temperatures.

Upon evaluating $\Delta(T)$, the thermodynamics of the system can be analyzed 
following Ref.\cite{bardeen}.  Let us start with the entropy:
\bea
S_{s} &=& -4\int_{0}^{\infty} dE \,N(E)(f\ln f+(1-f)\ln(1-f)),
\eea
where $f$ is the Fermi-Dirac distribution $(1+e^{\beta E})^{-1}$ with $\beta= 1/k_{B}T$.  N(E) 
is the quasiparticle density of states,
\bea
N(E)= N_{0} Re \langle \frac{|E|}{\sqrt{E^{2}-\Delta^{2} f^{2}}} \rangle
\eea
The electronic specific heat can be derived from the entropy via 
\bea
C_{s} = T\frac{\partial S}{\partial T}
\eea
In Fig. 5 we show $C_{s}/\gamma_{S}T$ versus $T/T_{c}$ for both phases.  Here $\gamma_{S} = 2\pi^{2}N_{0}/3$
is the Sommerfeld constant.  We find the jump $\Delta C/C$ at T$_c$ to be approximately 0.93 for the A-phase 
and 1.20 for the B-phase.  Data from Vollmer et al\cite{vollmer} shows these jumps to both be of order one, so that
our model is consistent with this data.  In addition, as expected, the low-temperature specific heat is predicted 
to be proportional to $T^{2}$\cite{maki-euro} for both phases.  Unfortunately, the presence of a 
Schottky specific heat peak\cite{vollmer} at low temperature makes assessment of the $T^{2}$ prediction difficult.

\begin{figure}
\resizebox{0.75\columnwidth}{!}{
\includegraphics{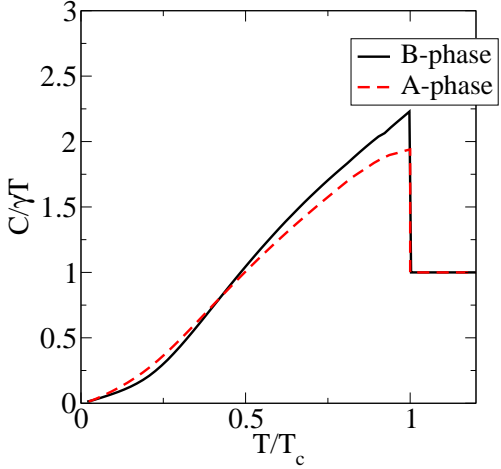}}
\caption{Specific heat C$_s$/$\gamma_{S}T$ for PrOs$_{4}$Sb$_{12}$}
\end{figure} 

Also the thermodynamical critical field $H_{c}(T)$ can be obtained from

\bea
F_{S}(T)-F_{N}(T) &=& -\int_{T}^{T_{c}}dT \,(S_{S}(T)-S_{N}(T)) \\
                  &=& -\frac{1}{8\pi}H_{c}^{2}(T)
\eea

Here S$_{S}$(T) and S$_{N}$ = $\gamma_{N}T$ are the entropies in the superconducting 
and normal state respectively.  We show $D(\frac{T}{T_{c}})= 
H_{c}(T)/H_{c}(0) - (1 - (T/T_{c})^{2})$ for both phases in Fig. 6.   The function 
\begin{figure}
\resizebox{0.75\columnwidth}{!}{
\includegraphics{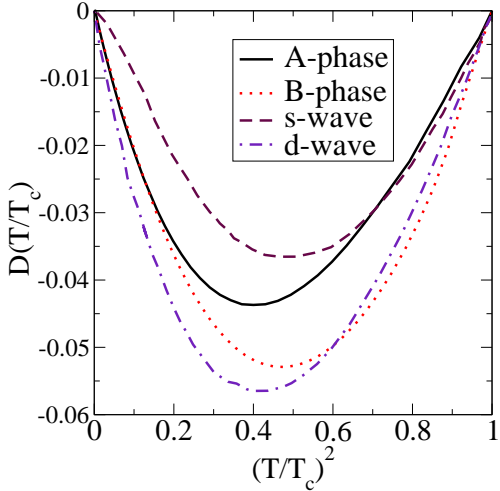}}
\caption{Deviation of critical field from parabolic dependence for p+h, s, and d-wave superconductors}
\end{figure}$D(\frac{T}{T_{c}})$ for both the A-phase and B-phase cases is slightly larger than for 
the isotropic s-wave case and somewhat smaller than for the d-wave case\cite{won94}, with the
B-phase case slightly larger than the A-phase case.
 
While there are a few reports of $H_{c2}(T)$ for the A-phase and $H^{*}(T)$, the phase boundary
between the A phase and the B phase\cite{bauer,izawa,measson,tenya,ho2,tayama}, no 
experimental data are available for $H_{c}(T)$.

Finally the superfluid density is given by
\bea
\frac{\rho_{s\parallel}(T)}{\rho_{s\parallel}(0)} = 1\, - 3 \int_{0}^{\infty}\frac{dE}{2T} \mathrm{sech}^{2}(E/2T) \times \nonumber \\
Re \left\langle z^{2} \frac{E}{\sqrt{E^{2}-\Delta^{2}(T)f^{2}}}\right\rangle
\eea
and 
\bea
\frac{\rho_{s\perp}(T)}{\rho_{s\perp}(0)} = 1 -\, \frac{3}{2}\int_{0}^{\infty}\frac{dE}{2T} \mathrm{sech}^{2}(E/2T) \times \nonumber \\
Re\left\langle (1-z^{2}) \frac{E}{\sqrt{E^{2}-\Delta^{2}(T)f^{2}}}\right\rangle
\eea
where Re$\langle \ldots \rangle$ refers to the real part, and the subscripts $\parallel$ and $\perp$ indicate parallel
and perpendicular directions to the nodal points.  The superfluid density, as expected, is isotropic
for the cubic symmetry-retaining A-phase, but rather anisotropic for the B-phase.  These superfluid densities are shown in Fig. 7.  In the low-temperature 
\begin{figure}
\resizebox{0.75\columnwidth}{!}{
\includegraphics{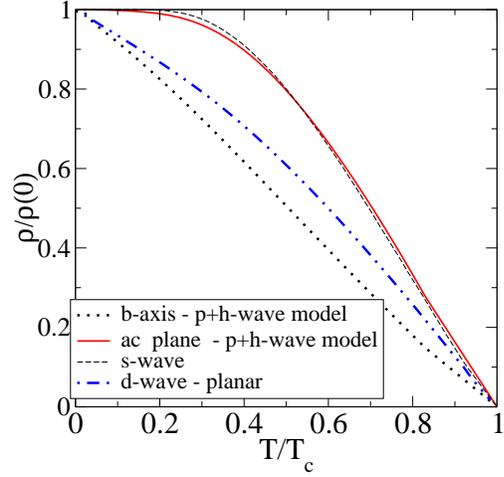}}
\caption{Predicted superfluid densities for PrOs$_{4}$Sb$_{12}$}
\end{figure}
regime ($T \ll \Delta$) 
both Eq.(20) and Eq.(21) can be expanded as 
\bea
\frac{\rho_{sA}(T)}{\rho_{s}(0)} &=&
1\, - \frac{\pi}{2}(\ln 2)\frac{T}{\Delta}+\ldots \\
\frac{\rho_{sB\parallel}(T)}{\rho_{s\parallel}(0)} & = & 
1\, - \frac{3\pi}{4}(\ln 2)\frac{T}{\Delta} + \ldots \\
\frac{\rho_{sB\perp}(T)}{\rho_{s\perp}(0)}& = & 1 -\,\frac{\pi^{2}}{16}(\frac{T}{\Delta})^{2} + \ldots\,\,\,\,.
\eea
\begin{figure}
\resizebox{0.75\columnwidth}{!}{
\includegraphics{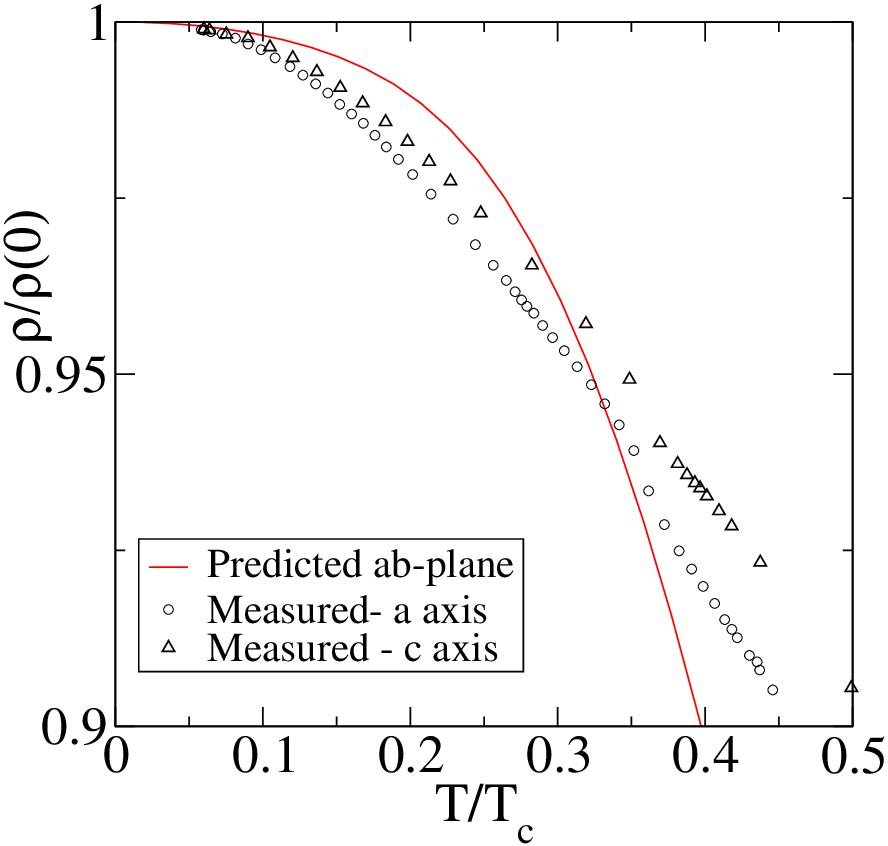}}
\caption{Superfluid densities for PrOs$_{4}$Sb$_{12}$}
\end{figure}Close to the transition temperature, we find
\bea
\frac{\rho_{sA}(T)}{\rho_{sA}(0)} & \simeq & 
\frac{6}{7}\frac{<f^{2}>}{<f^{4}>}(-\ln T/T_{c}) \\
&=& 1.393 (-\ln T/T_{c})\\
\frac{\rho_{sB\parallel}(T)}{\rho_{s\parallel}(0)} & \simeq  & 
\frac{17 \cdot 13}{21 \cdot 11}(-\ln T/T_{c}) \\
&=& 0.9567(-\ln T/T_{c}) \\
\frac{\rho_{sB\perp}(T)}{\rho_{s\perp}(0)} & \simeq & 
\frac{31\cdot 221}{45 \cdot 77}(-\ln T/T_{c}) \\
&=& 1.9772 (-\ln T/T_{c})
\eea
In the figure above we have compared $\frac{\rho_{sB\parallel}(T)}{\rho_{sB\parallel}(0)}$ with the
data taken from Chia et al\cite{chia}, assuming that the nodal points
in $\Delta({\bf k})$ are aligned parallel to ${\bf H}$.  Rather satisfactory
agreement is observed for $T < T_{c}/3$.  But the theoretical $\rho_{s\perp}(T)$ vanishes linearly 
with $T_{c}-T$ in the vicinity of $T=T_c$, whereas Chia et al \cite{chia} found a $\rho_{s\perp}(T)$
which vanishes with essentially infinite slope at $T_{c}$. 

Recently Chia et al \cite{chia-2} also reported magnetic penetration depth measurements for a range
of dopings x from 0.1 to 0.8 in the compound Pr(Os$_{1-x}$Ru$_{x}$)$_{4}$Sb$_{12}$.  Over the range from $x=0.4$ to
$x=0.8$, exponential temperature dependence of the superfluid density was found, indicating an isotropic 
s-wave gap function in this regime.  Of direct interest for this work, the superfluid density was 
found to go to zero linearly for all dopings, with no hint of the essentially infinite slope found
\cite{chia} in the pure case (x=0).  In addition, the slope of these linear curves at T$_{c}$ 
does not increase dramatically from x=0.8 to x=0.1.  Further experiments at doping ranges between $x=0$ and $x=0.1$ are highly desirable, to examine more closely the apparent transition from nodal to conventional superconductivity
taking place in this system.  It would also be of value to confirm the rather unusual ``infinite-slope'' 
behavior observed in the pure sample near T$_{c}$.

\section{Concluding Remarks}

We have worked out the weak-coupling theory of the 
A\, and\, B\, phases\, of\, the\, heavy-fermion\, superconductor \\
PrOs$_{4}$Sb$_{12}$.  A simple thermodynamic analysis offers an explanation
for the appearance of the lower-symmetric B phase at lower temperatures.  
The present model leads to a fair description of STM data taken by Suderow 
et al \cite{suderow}.
In addition, the present model for the B-phase
describes the superfluid density determined by Chia et al\cite{chia} for the low-temperature
regime, if we assume that the nodal points in the B-phase follow the magnetic field direction
in the field cooled situation\cite{maki1}.  Since the magnetic field is the only symmetry-breaking
agent, this appears to be plausible.  We will present the results of an analysis in the case
of impurities shortly.

\small{We thank H. Won, P. Thalmeier, K. Izawa, Y. Matsuda and H. Tou for many useful discussions.
S.H. acknowledges financial support through PRF grant 41757-AC10.}


\begin{thebibliography}{}
\bibitem{footnote} Therefore a few of the B-phase
results presented here are the same as in the analysis performed in Ref.\cite{maki-euro}; 
we include them here for completeness.
\end{thebibliography}
\end{document}